\documentclass[12pt]{article}
\usepackage{graphics,geometry,epsfig,pslatex}
\usepackage{amsmath,amssymb}
\usepackage[dvips]{color}
\usepackage{times, multicol}

\newcommand{\mb}{\mathbf}

\newcommand{\ec}{\mbox{e}}
\newcommand{\ic}{\mbox{i}}

\newcommand{\CmtIp}[4]{\langle \mb{E}_{#1}^{#2}, \mb{H}_{#1}^{#2} ; \mb{E}_{#3}^{#4},  \mb{H}_{#3}^{#4} \rangle}

\newcommand{\mum}{\,\mu\mbox{m}}
\newcommand{\TE}[1]{TE$_{#1}$} % mode indexing
\newcommand{\mbss}[1]{\mbox{\scriptsize #1}}

\newcommand{\mbsf}[1]{\mbox{\sf #1}}

\begin{document}
% --- title
\begin{center}
{\LARGE\bf
Multimode circular integrated optical micro- resonators: Coupled mode theory modeling
}\\[3mm]
% --- author
K.~R.~Hiremath, R.~Stoffer, M.~Hammer \\ 
email: k.r.hiremath@math.utwente.nl
\\[2mm]
% --- affiliation
{\footnotesize
  MESA$^{+}$ Research Institute, University of Twente, The Netherlands
}
\end{center}

% --- abstract
{\it 
  A frequency domain model of multimode circular microresonators for filter applications in integrated optics is investigated. Analytical basis modes of 2D bent waveguides or curved interfaces are combined with modes of straight channels in a spatial coupled mode theory framework. Free of fitting parameters, the model allows to predict quite efficiently the spectral response of the microresonators. It turns out to be sufficient to take only a few dominant cavity modes into account. Comparisons of these simulations with computationally more expensive rigorous numerical calculations show a satisfactory agreement. 
}

%-------------------------------------------------------------------------
\vspace{-0.5cm}
\subsection*{Introduction}
\begin{figure}[htb]
\vspace{-0.5cm}
\begin{minipage}{0.5\linewidth}
 \epsfig{file=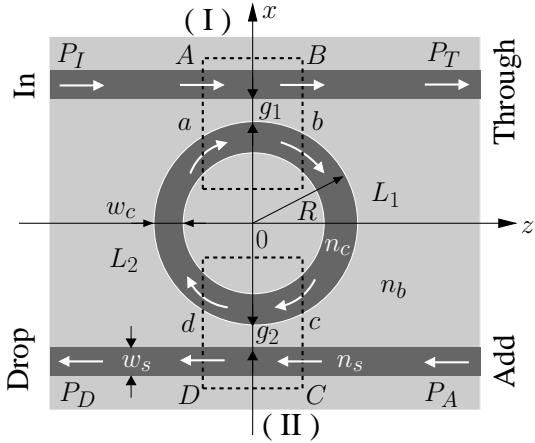, width=\linewidth}
\end{minipage}
\begin{minipage}{0.48\linewidth}
  \caption{\footnotesize  Functional decomposition of 2D microresonators: $R$
    is the radius of the cavity with a core refractive index $n_c$ and a width
    $w_c$ ($=0$, for disks). The straight waveguides have a core refractive
    index $n_s$ and a width $w_s$; $g_1$ and $g_2$ are the separation distances between the cavity and
  the straight waveguides; $n_{b}$ is the background refractive index; $L_{1}$
    and $L_{2}$ denote the lengths of the cavity segments which are not
    included in the coupler regions. The letters $A, B, C, D$ and $a, b, c, d$ denote the coupler port planes.}
  \label{2D_MR}
\end{minipage}
\end{figure}

Nowadays due to their superior selectivity, compactness, and possibility of
dense integration, microresonators (MRs) become attractive for application as
wavelength add/drop filters \cite{NAIS}. A typical microresonator setting, where a ring/disk shaped cavity is
placed between two straight waveguides, is shown in Fig.~\ref{2D_MR}. In this
paper, we outline a spatial coupled mode theory (CMT) based model of 2D
circular  multimode optical microresonators.
\vspace{-0.7cm}

%--------------------------------------------------------------------------
\subsection*{Abstract resonator model}
For modeling purposes, the MR is decomposed into two bent-straight waveguide
couplers, represented by the blocks (I) and (II), which are connected to each other by two segments of the cavity. The external
connections are provided by straight waveguides. Here we consider only forward
propagating modes of uniform polarization. We assume that all elements are
linear and that the back reflections inside the couplers and the cavity are
negligible. Outside the couplers, the interaction between the constituent
waveguides is assumed to vanish. We consider a frequency domain description of the optical field. The vacuum
wavelength $\lambda$ prescribes the real angular frequency $\omega$. Assume
that $N_{s}$ modes of the straight waveguides and $N_{b}$ bend modes of the cavity are taken into account. 

Let $\gamma^{p}$ be the complex propagation constant of the $p$'th cavity mode. The variables $A^{q}, B^{q}, C^{q}, D^{q}$  and $a^{p}, b^{p}, c^{p}, d^{p}$
denote the directional amplitudes of these properly normalized modes in the coupler port planes, combined into amplitude vectors $\mb{A}, \mb{B}, \mb{C}, \mb{D}$ and $\mb{a}, \mb{b}, \mb{c}, \mb{d}$. Let $\mbsf{S}^{\mbss{I}}$ and $\mbsf{S}^{\mbss{II}}$ be the
scattering matrices for coupler I and  II respectively, i.e.
\vspace{-0.2cm}
\begin{equation} 
\begin{bmatrix} \mb{b} \\ \mb{B} \end{bmatrix} = \mbsf{S}^{\mbss{I}}
\begin{bmatrix} \mb{a} \\ \mb{A} \end{bmatrix},
\hspace{1cm}
\begin{bmatrix} \mb{d} \\ \mb{D} \end{bmatrix} = 
\mbsf{S}^{\mbss{II}}
\begin{bmatrix} \mb{c} \\ \mb{C} \end{bmatrix}.
\label{amp1}
\vspace{-0.1cm}
\end{equation}
The amplitudes of the connecting cavity segments are related to each other as
\begin{equation}
c^{p}  =  b^{p} \ec^{(- \ic \gamma^{p} L_{1})}, \hspace{0.5cm}     
a^{p}  =  d^{p} \ec^{(- \ic \gamma^{p} L_{2})}.  
\label{amp2}
\vspace{-0.1cm}
\end{equation}
Given input powers $P^{q}_{\mbss{I}} = |A^{q}|^2$ at $A$ and
$P^{q}_{\mbss{A}} = |C^{q}|^2$ at $C$ , we are interested in the transmitted powers
$P^{q}_{\mbss{T}} = |B^{q}|^2$ at $B$ and the backward dropped powers
$P_{\mbss{D}}^{q} = |D^{q}|^2$ at $D$. This means solving the linear system of
equations \eqref{amp1},~\eqref{amp2} for $B^{q}$ and $D^{q}$, for $q=1, 2,
\ldots, N_{s}$. When scanned over a wavelength range, resonances appear as maxima of the dropped power and minima of the transmitted power. 

To evaluate the microresonator model described above, one must know the cavity
propagation constants $\gamma^{p}$ and the scattering matrices
$\mbsf{S}^{\mbss{I}}$, $\mbsf{S}^{\mbss{II}}$. Using an analytic model of bent
waveguides as described  in Ref.~\cite{HHS041}, we obtain the bend modes and
their propagation constants. Having access to the bend modes, a model of the
bent-straight waveguide couplers in terms of CMT leads to the scattering
matrices.  A detailed description of this procedure for one cavity mode and one
straight waveguide mode is presented in Ref.~\cite{HHS04}. In the next
sections, we extend it to the case of multimode MRs.

%--------------------------------------------------------------------------
\subsection*{Multimode bent-straight waveguide couplers}
\begin{figure}[htbp]
\begin{minipage}{0.55\linewidth}
  \centering
  \epsfig{file=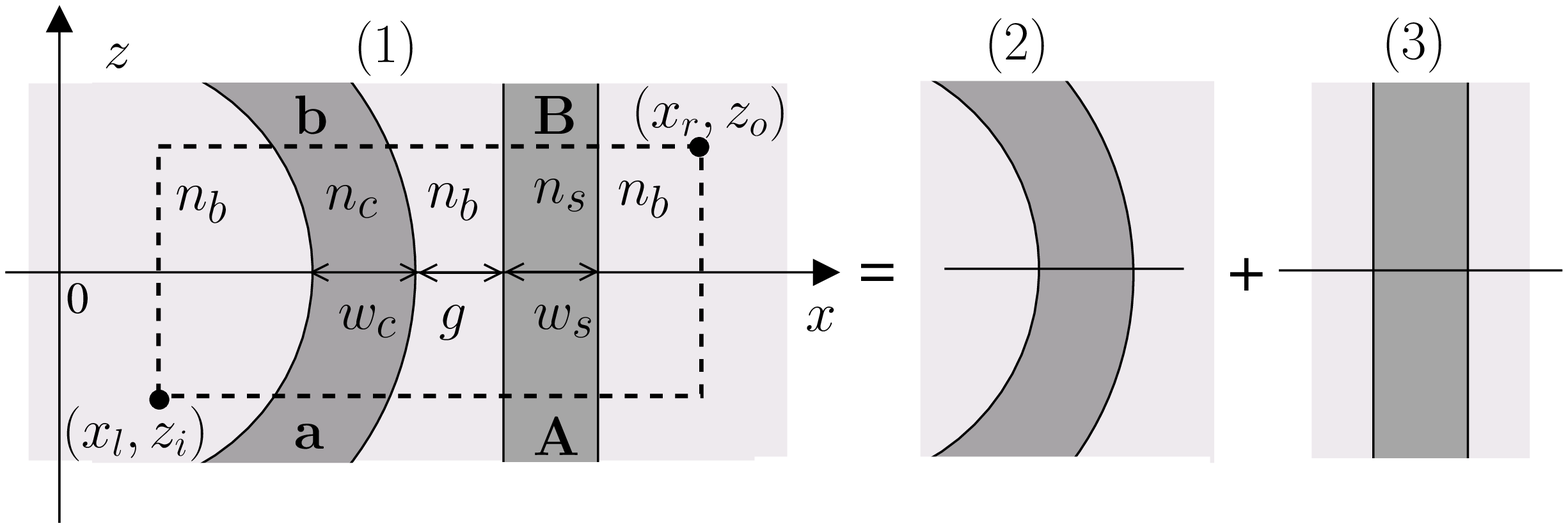, width=\linewidth}
\end{minipage}
\begin{minipage}{0.44\linewidth}
\vspace{-0.5cm}
  \caption{\footnotesize CMT setting for bent-straight waveguide couplers. The coupler is defined in the region $[x_{l}, x_{r}]
\times [ z_{i}, z_{o}]$. The external segments of the bent waveguide and the
straight waveguide constitute the port connections. It is assumed that outside the coupler there is negligible interaction between the waveguides. 
}
  \label{BSC_cmt}
\end{minipage}
\end{figure}

Consider the bent-straight waveguide coupler as shown in
Fig.~\ref{BSC_cmt}-(1). Let $\{\mb{E}_{b}^{p}, \mb{H}_{b}^{p}, \epsilon_{b} \} $ and
 $\{\mb{E}_{s}^{q}, \mb{H}_{s}^{q}, \epsilon_{s} \} $ represent the modal
 electric fields, magnetic fields, and the spatial distributions of the
 relative  permittivity of the bent waveguide and the straight waveguide
 respectively. Here the modal fields include the harmonic dependence on the propagation coordinate and are expressed in the Cartesian coordinates $(x, z)$.
\vspace{-0.4cm}
\vspace{-0.3cm}
  The field $\{\mb{E}, \mb{H} \}$ inside the coupler is  given by a linear combination of the modal fields of the bent waveguide (Fig.~\ref{BSC_cmt}-(2)) and the  modal fields of the straight waveguide (Fig.~\ref{BSC_cmt}-(3)):
\begin{equation}
  \label{cmt_ansatz}
\vspace{-0.2cm}
\begin{bmatrix} \mb{E}(x,z) \\ \mb{H}(x,z) \end{bmatrix} =
\sum_{p=1}^{N_{b}} C_{b}^{p}(z) \begin{bmatrix} \mb{E}_{b}^{p}(x,z) \\ \mb{H}_{b}^{p}(x,z) \end{bmatrix} + 
\sum_{q=1}^{N_{s}} C_{s}^{q}(z) \begin{bmatrix} \mb{E}_{s}^{q}(x,z) \\ \mb{H}_{s}^{q}(x,z) \end{bmatrix}, 
\end{equation}
where $C_{b}^{p}$, $C_{s}^{q}$ are unknown amplitudes. As shown in Ref.~\cite{SHH04}, a procedure based on the Lorentz reciprocity theorem leads to the following coupled mode equations 
\vspace{-0.2cm}
\begin{equation}
\mbsf{M}(x,z) \cdot d_{z} \mb{C}(z) = \mbsf{F}(x,z) \cdot \mb{C}(z), \label{CMEs}
\vspace{-0.2cm}
\end{equation}
with $\mbsf{M}_{ij} = \CmtIp{m}{i}{n}{j} =  \int \mb{a}_z \cdot (\mb{E}_{m}^{i} \times \mb{H}^{j*}_{n} + \mb{E}^{j*}_{n} \times \mb{H}_{m})  dx$, $\mbsf{F}_{ij} = \int (\epsilon - \epsilon_{m}) \mb{E}_{m}^{i} \cdot \mb{E}_{n}^{j*} dx$, $C_{i}  = C_{m}^{i}$ for $i,j=1, 2,
 \cdots, (N_b + N_s)$, and $m= b$ if $ 1 \leq i \leq N_{b}$ otherwise $m=s$, $n= b$
 if $ 1 \leq j \leq N_{b}$ otherwise  $n=s$. Here $\mb{a}_{z}$ is a unit vector in
 $z$- direction and $\epsilon$ is the relative permittivity of the complete
 coupler. The integrations extend over $[x_{l}, x_{r}]$ for each $z$ level.

Solving Eq.~\eqref{CMEs} by a Runge Kutta method of order 4, we get a
relation  $\mb{C}(z_{o}) = \mbsf{T} \cdot \mb{C}(z_{i})$ between the
amplitudes in the output and input coupler ports. For rather radiative
bend modes, it takes a long $z$- distance to stabilize the elements of matrix
$\mbsf{T}$. This difficulty is overcome by taking the projections of the
coupled fields onto the straight waveguide modes \cite{SHH04}. Then the output
amplitudes of the straight waveguide modes are given by  \vspace{-0.3cm}
\begin{equation}
  \label{proj_amp}
  B^{q} = \hspace{-0.1cm} \left[  C_{s}^{q}(z_o) + \hspace{-0.1cm} \left .\sum_{p=1}^{N_b} C_{b}^{p}(z_o) \frac{\CmtIp{b}{p}{s}{q}}{\CmtIp{s}{q}{s}{q}}  \right |_{z_{o}} \right] \ec^{-\ic \beta^{q} z_{o}}.
\end{equation}
By incorporating these projection corrections into $\mbsf{T}$, we finally
obtain  the required scattering matrix $\mbsf{S}$. 

%--------------------------------------------------------------------------
\vspace{-0.2cm}
\subsection*{Simulations and comparison}
We consider a cavity in the form of a disk, i.e. $w_{c}=0$. Since the modal
loss of the bend modes increases with growing mode order (where the order of the mode
is defined as in Ref.~\cite{HHS041}), one can expect that only the lower order
bend modes play a dominant role for the field evolution in the cavity. 
\begin{figure}[!htb]
\vspace{-0.2cm}
  \centering
\epsfig{file=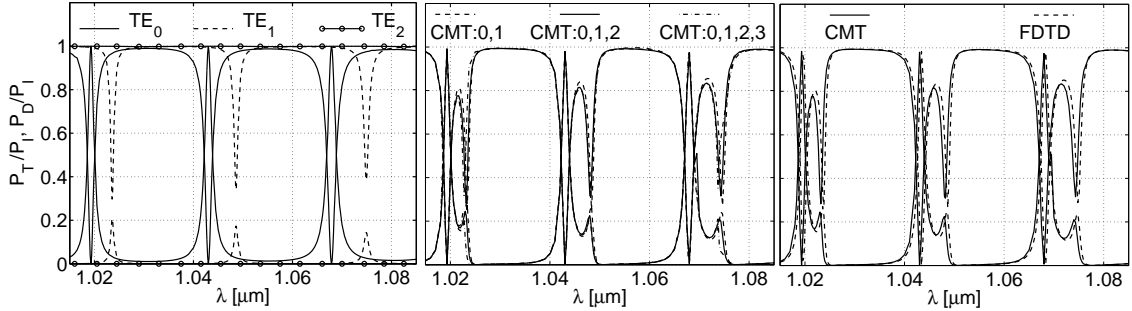, width=\linewidth}
\vspace{-0.8cm}
  \caption{\footnotesize TE power spectrum for a 2D disk microresonator. \underline{Left}:
    CMT with only a single (one of \TE{0},\TE{1} or \TE{2}) cavity
  mode. \underline{Middle}:  CMT results for two (dashed line), three (solid line) and four (dash-dotted line) cavity modes together. \underline{Right}: Comparison of FDTD results and CMT results with 3 cavity modes. MR
  specifications: $n_{c}=n_{s}=1.5$, $n_{b}=1.0$, $w_{c}=0 \ \mum$ (disk), $w_{s}=0.4 \ \mum$, $R=5 \ \mum$,  $g=0.2 \ \mum$. }
  \label{cmt_fdtd_ps}
\end{figure}

The present CMT setting allows to investigate the significance of individual
cavity modes for the spectral response of the MRs. The left plot of
Fig.~\ref{cmt_fdtd_ps} shows the dropped power and the transmitted power when
only a single cavity mode (either \TE{0}, \TE{1} or \TE{2}) is included
in the CMT model. The extrema corresponding to the fundamental mode (\TE{0})
are much more pronounced than those related to the first order mode (\TE{1}). If only the \TE{2} cavity mode is
taken into account, on the present scale hardly any variations of $P_{D}$ and
$P_{T}$ appear. The middle plot of Fig.~\ref{cmt_fdtd_ps} shows the cumulative
effect of the higher order cavity modes. The extrema corresponding to the fundamental mode remain almost unaffected,
but the shape of the resonances related to the \TE{1} mode changes. Note that the
curves  for three and four cavity modes almost coincide. The right plot of
Fig.~\ref{cmt_fdtd_ps} compares the CMT results (3 cavity modes) with FDTD
simulations \cite{Sto01}.  The results agree surprisingly well.

\begin{figure}[!h]
 \vspace{-0.1cm}
  \centering
  \epsfig{file=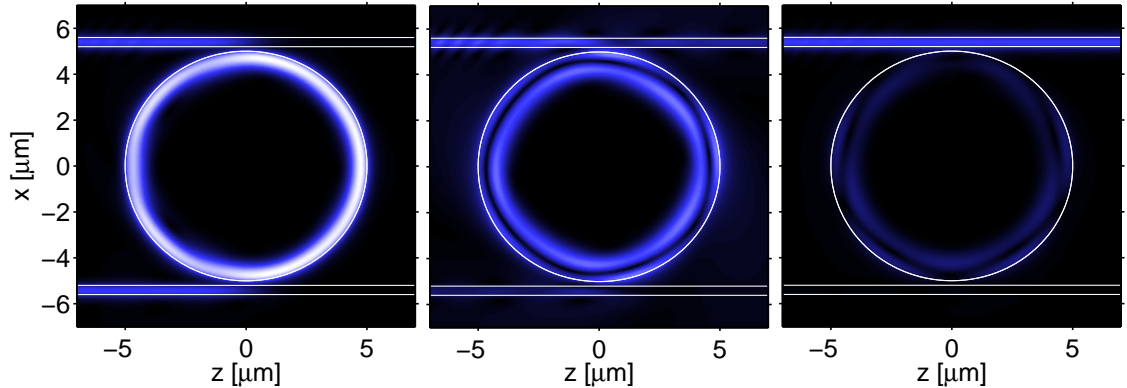, width=\linewidth} 
  \vspace{-0.8cm}
  \caption{\footnotesize Absolute value of the $y$ component of TE fields of
  the MR. The gray scales (black = zero) are comparable among the plots.  From
  left to right: MR field at a resonance
  corresponding to the fundamental mode ($\lambda=1.043 \mum$),  a resonance
  corresponding to the higher modes ($\lambda=1.04833 \mum$) and  off
  resonance ($\lambda=1.055 \mum$). CMT simulations ( 3 cavity modes) for a MR
  with the parameters as for Fig.~\ref{cmt_fdtd_ps}.}
  \label{mr_plots}
\end{figure}

Fig.~\ref{mr_plots} illustrates the field profiles for the full MR structure as
predicted by the CMT. At the resonance corresponding to the fundamental mode, most
of the input power is coupled to the fundamental cavity mode and appears at
the drop port. For the resonance corresponding to the higher order  modes, the
circular nodal line in the field pattern in the cavity indicates that a
significant part of the input power is coupled to the \TE{1} mode. For an off
resonance wavelength, most of the input power appears at the through port.
%-----------------------------------------------------------------------
\vspace{-0.8cm}
\subsection*{Conclusions}
The CMT based model of 2D microresonators yields quite accurate results even if
few cavity modes are taken into account. These results agree  well with the
rigorous FDTD results and can be obtained with much lower computational effort. The role of the individual cavity modes can be clearly identified in this CMT model. 
%-----------------------------------------------------------------------

\vspace{-0.4cm}
\subsection*{Acknowledgment}
\vspace{-0.1cm}
This work has been supported by the European
  Commission (project IST-2000-28018, `NAIS'). The authors thank  E. van
  Groesen and H.~J.~W.~M.~Hoekstra  for many fruitful discussions on the subject. 
%-------------------------------------------------------------------------
\vspace{-0.3cm}
%{\footnotesize
%\bibliographystyle{unsrt}
%\bibliography{waveguide,microresonator,mathematics}
%}

{\footnotesize
\bibliographystyle{unsrt}

\vspace{-0.2cm}
Cf. also the bibliography lists of Refs.[2]-[5].
}
%-------------------------------------------------------------------------

%% \bibliographystyle{unsrt}
%% \begin{thebibliography}{9}\setlength{\itemsep}{-3pt}\footnotesize
%%   \bibitem{ref1} A.B. Smith and D.E. Jones, "Title of paper", {\it Journal
%%     name}, vol. 10, pp. 1832-1838, 1992.
%%   \bibitem{ref2} J. Jansen and J. Janssens, Jr., "Title of paper", in {\it
%%     Proceedings of the Conference ...}, 1996, pp. 123-125.
%%   \bibitem{ref3} A.B. Author, Title of Book, New York: IEEE Press, 1995,
%%     ch. 6, pp. 23-35.
%%   \bibitem{ref4} P.Q. Roberts, "Title of chapter in book", in {\it Title of
%%     Book}, R.E. Dak and T. Euren, Eds., Am\-sterdam: North-Holland, 1991, pp.
%%     123-456.
%% \end{thebibliography}

\end{document}